\documentclass[
preprint, 
superscriptaddress,
 amsmath,amssymb,
 aps,
floatfix,
pre,
]{revtex4-1}

\usepackage{graphicx}
\usepackage{dcolumn}
\usepackage{bm}

\usepackage{color}
\newcommand{\skb}[1]{\textcolor{black}{#1}}

\begin{document}

\title{Threshold behavior of a social norm in response to error proneness}%

\author{Quang Anh Le}
\affiliation{Industry-University Cooperation Foundation, Pukyong National
University, Busan 48513, South Korea}
\author{Seung Ki Baek}
\email{seungki@pknu.ac.kr}
\affiliation{Department of Scientific Computing, Pukyong National University,
Busan 48513, South Korea}
\date{\today}

\begin{abstract}
A social norm defines what is good and what is bad in social contexts, as well as what to do based on such assessments. A stable social norm should be maintained against errors committed by its players. In addition, individuals may have different probabilities of errors in following the norm, and a social norm would be unstable if it benefited those who do not follow the norm carefully. In this work, we show that Simple Standing, which has been known to resist errors and mutants successfully, actually exhibits threshold behavior.
That is, in a population of individuals playing the donation game according to Simple Standing, the residents can suppress the invasion of mutants with higher error proneness only if the residents' own error proneness is sufficiently low. Otherwise, the population will be invaded by mutants that commit assessment errors more frequently, and a series of such invasions will eventually undermine the existing social norm.
This study suggests that the stability analysis of a social norm may have a different picture if the probability of error itself is regarded as an individual attribute.
\end{abstract}

\maketitle


\section{Introduction}

To err is human. A theory of human behavior should take into account the possibility of error in order to be a realistic description~\cite{myerson2013game}.
Indirect reciprocity is a theory that explains reputation-based cooperation in a large human population~\cite{nowak1998evolution,nowak2005evolution}.
Its performance heavily relies on how precisely one can process the information of others, but our understanding of error has often been minimalist in the sense that it is usually assumed to be an environmental factor, which is inevitable and uncontrollable to a large extent from an individual's point of view.
Although this assumption contains an important piece of truth, if we think of our daily experience, it also seems entirely plausible that some are more prone to errors than others.
In this sense, it has been largely overlooked that the error could, at least partly, be an individual attribute.
\skb{We will consider errors in assessing each other throughout this work. Note that assessment errors should be distinguished from perception errors. If an observer is an unconditional cooperator, a perception error does not change the observer's assessment, while the assessment error does.}

In a previous work~\cite{le2025response}, we assumed that each individual might have a different probability of \skb{assessment} error in observing a social norm.
Each individual's probability of \skb{assessment} error was regarded as a quenched random variable, which does not evolve in time, and the question was how disadvantageous error proneness would be.
The social norms studied were the `leading eight'~\cite{ohtsuki2004should,ohtsuki2006leading,fujimoto2022reputation,fujimoto2023evolutionary}, and we numerically found that Simple Standing does not impose disadvantages on those who are careless in observing the norm. This finding needs explanation because Simple Standing has been known to successfully resist error and mutation among the leading eight~\cite{lee2022second,fujimoto2024leader}.

In this work, we consider this problem in a simpler setting by borrowing a framework from mutation-selection dynamics. We begin by considering a population with a certain probability of assessment error $\epsilon_1$, and individual heterogeneity is introduced when an infinitesimally small fraction of mutants appear with a different probability $\epsilon_0$.
The crucial question is whether such a mutant earns a higher average payoff than a resident player. Our analytic and numerical calculations reveal how the robustness of Simple Standing can be reconciled with its favorable response to error proneness reported in Ref.~\onlinecite{le2025response}. The point is that Simple Standing with $\epsilon_1$ exhibits threshold behavior in the sense that a resident player is better off than an error-prone mutant only when $\epsilon_1$ is below a certain threshold. Above the threshold, it ceases to suppress the mutant, and this is consistent with the observation in Ref.~\onlinecite{le2025response}, where individuals were allowed to have any error probabilities between $0$ and $1/2$.

Although we use the terminology of mutation and selection, it is worth stressing that the mutant still shares the same norm as the residents and that only the probability of assessment errors differs. We are thus looking at the collapse of a social norm \emph{from within}, rather than an exogenous replacement by another competing social norm as considered in the literature~\cite{fujimoto2024leader,murase2024computational}.
Cultural transmission is a complicated process that may not always be equated with fitness-governed genetic inheritance~\cite{kim2021win}, and a part of the reason is that a social norm can protect itself from external forces by mutual reinforcement between expectation and action~\cite{mackie2015social}. Therefore, instead of touching on the dynamics of two competing norms, here we wish to consider one of its preconditions, as has been expressed by historians~\cite{durant1944story} when they say, ``A great civilization is not conquered from without until it has destroyed itself within.''

This work is organized as follows. The next section explains our simulation method and the analytic solutions.
We will perform numerical calculations with the conventional discrete model and compare the results with those obtained by solving the continuous model~\cite{lee2021local,lee2022second,mun2023second}.
The results will be presented in Sec.~\ref{sec:results}, where we can see the agreement between the numerical and analytic approaches. We discuss the results and summarize this work in Sec.~\ref{sec:summary}.

\section{Method}

Let us start by explaining the discrete model. Imagine a large population of size $N$.
Each individual $k$ has an opinion $m_{ki}^t \in \{0,1\}$ about another individual $i$ at time $t$. We have $m_{ki}^t=1$ when $k$ has a good opinion about $i$ and $m_{ki}^t=0$ when it is a bad opinion. At each time step, two random individuals $i$ and $j$ are chosen for a donation game, parametrized by the benefit $b$ and the cost $c$ of cooperation with $b>c$. The former individual $i$, playing the role of the donor, determines how much to donate to $j$ \skb{by referring to his or her own behavioral rule $\beta_i$}.
\skb{We assume that every individual is a discriminator when playing the role of a donor, so that the donor will cooperate if and only if the recipient looks good to the donor. In other words, the donor $i$'s action toward the recipient $j$ at time $t$ is written as $\beta_{ij}^t = \beta \left(m_{ij}^t \right) = m_{ij}^t$, where $\beta_{ij}^t=1$ and $0$ mean cooperation and defection, respectively.
}

The individual $k$ observes $i$'s decision and updates his or her own opinion about the donor $i$ according to an assessment rule $\alpha_k$, which is a function of $k$'s opinions about $i$ and $j$ and of $i$'s behavior toward $j$.
However, with probability $\epsilon_k$, the assessment $\alpha_k$ can change from $0$ to $1$ or the other way around. If $\alpha_k = 1$, the observer $k$'s opinion about the donor $i$ is updated to good, and $\alpha_k=0$ means that it is updated to bad.
\skb{We assume that everyone shares a common assessment rule such that $\alpha_k = \alpha$ for every $k$.
A mutant has index $k=0$ with a probability of error $\epsilon_0$. All others with indices $k>0$ are residents with a common probability of error $\epsilon_1 = \epsilon_2 = \ldots = \epsilon_{N-1}$. An individual with $k=1$ can thus be a representative of the resident population.
If $\pi_k$ denotes $k$'s average payoff per round in the long-time limit,
our primary interest is in the relative payoff advantage of a mutant,
\begin{equation}
\Delta \pi_0 = b m_{10} - c m_{01} - (b-c) m_{11},
\label{eq:ss_dpi0}
\end{equation}
where we have suppressed the time index as $t \to \infty$.
}

\begin{table}
\caption{Four social norms considered in this work. Throughout this work, we map `cooperation ($C$)' and `good ($G$)' to $+1$, and `defection ($D$)' and `bad ($B$)' to $0$.}
\begin{ruledtabular}
\begin{tabular}{c|cccc|cc}
 & $\alpha(C,G)$ & $\alpha(D,G)$ & $\alpha(C,B)$ & $\alpha(D,B)$ &
 $\beta(G)$ & $\beta(B)$\\\hline
Image Scoring & $G$ & $B$ & $G$ & $B$ & $C$ & $D$\\
Stern Judging & $G$ & $B$ & $B$ & $G$ & $C$ & $D$\\
Shunning & $G$ & $B$ & $B$ & $B$ & $C$ & $D$\\
Simple Standing & $G$ & $B$ & $G$ & $G$ & $C$ & $D$
\end{tabular}
\label{tab:norms}
\end{ruledtabular}
\end{table}

We restrict ourselves to Image Scoring, Stern Judging, Shunning, and Simple Standing, which are second-order norms that have been widely studied in the literature~\cite{sasaki2017evolution,okada2018solution,xia2020effect,radzvilavicius2021adherence,kessinger2023evolution}.
These are second-order norms in the sense that an observer should refer to the donor's action toward the recipient and the observer's opinion about the recipient when assessing the donor \skb{(see also Appendix~\ref{app:second} for a comprehensive analysis of all second-order norms)}. According to our notation, it means that $\alpha = \alpha \left(\beta_{ij}^t, m_{kj}^t \right)$.
The definitions of the four norms are given in Table~\ref{tab:norms}.

\subsection{Numerical approach}

\subsubsection{\skb{Payoff calculation}}

For agent-based model (ABM) simulations,
we use the conventional discrete model as described above. We start from a random initial configuration in which the $N \times N$ image matrix at $t=0$ is filled with zeros and ones equally probably. At each time step, we randomly pick up a donor $i$ and a recipient $j$. The donor chooses what to do to the recipient by calculating $\beta \left(m_{ij}^t \right) \in \{0,1\}$.
We then sample observers $N$ times with replacement. If the observer is
the donor or the recipient, the interaction is observed with certainty, while the other individuals observe the interaction with probability $q$. An observer $k$ updates his or her opinion about the donor by calculating the assessment function $\alpha \in \{0, 1\}$.
In the case of a second-order norm, $\alpha$ depends only on the donor's behavior and the observer's opinion about the recipient, so we may write $\alpha = \alpha \left( \beta_{ij}^t, m_{kj}^t \right)$.
After $T$ time steps when the system has reached a stationary state ($T \gg N$), we record the average level of donation of the population, together with how much the mutant individual with $i=0$ donates and receives from others on average. This procedure is repeated to obtain $S$ independent samples.

\subsubsection{\skb{Selection and mutation}}
\label{sec:selection}

\skb{When incorporating selection into the ABM simulations, we allow every individual $i$ to update his or her own $\epsilon_i$ at every $\tau_s$ time steps.
Specifically, we let each individual choose a random person $k$ as a mentor and adopt the mentor's $\epsilon_k$, if and only if the mentor has earned a higher payoff. To keep track of the time evolution of the error probabilities, we record the mean and standard deviation of $\left\{ \epsilon_i \right\}$, which are denoted as $\bar\epsilon$ and $\Delta\epsilon$, respectively. After each update, everyone's payoff is reset to zero. The period $\tau_s$ must be sufficiently long compared to the population size $N$ to observe the effect of assessment error statistically.}

\skb{Concerning mutation, we assume that each $\epsilon_i$ changes by a small amount $\delta$ at every $\tau_m$ time steps. This time scale of mutation should be sufficiently longer than $\tau_u$ for the population to have a well-defined average all the time.
The change $\delta$ is sampled from the Gaussian distribution with a standard deviation $w \ll 1$. If $\epsilon_i+\delta$ lies outside of $[0,1/2)$, the attempt is rejected.}

\subsection{Analytic approach}
In the continuous model of indirect reciprocity, $m_{ki}^t$ has a real value
$\in [0,1]$. The degree of cooperation also takes a continuous value, so $i$
donates $b\beta\left(m_{ij}^t\right)$ at the cost of
$c\beta\left(m_{ij}^t\right)$. The observer $k$'s opinion about the donor $i$
will be updated gradually by referring to another continuous function $\alpha$
used to evaluate the donor's action. The functions $\alpha$ and $\beta$ are
obtained from bilinear and linear interpolations of the conventional discrete norms~\cite{mun2023second}.
On average, the dynamics is thus governed by the following ordinary differential equation:
\begin{eqnarray}
m_{ki}^{t+1} &= (1-q) m_{ki}^t +& \frac{q}{N} \sum_{j=1}^N (1-\epsilon_{k}) \alpha_k \left[ \beta_i(m_{ij}^t), m_{kj}^t \right]\nonumber\\
&&+ \epsilon_{k} \left\{ 1- \alpha_k \left[\beta_i(m_{ij}^t), m_{kj}^t \right]  \right\}\label{eq:motion}\\
&= (1-q) m_{ki}^t +& q\epsilon_k + \frac{q}{N} \sum_{j=1}^N \left(1 - 2\epsilon_k\right) \alpha \left[ \beta \left( m_{ij}^t \right), m_{kj}^t \right],
\label{eq:motion2}
\end{eqnarray}
where $q$ is the probability of observation. Its specific value is irrelevant when we analyze a stationary state because it only changes the overall time scale. \skb{The first term in Eq.~\eqref{eq:motion} means that the existing value is maintained if the interaction is not observed. If the interaction is observed with probability $q$, the observer updates the assessment correctly according to $\alpha_k$ with probability $1-\epsilon_k$, or incorrectly with probability $\epsilon_k$. The factor of $1/N$ in front of the summation means that we are averaging the assessment of the donor $i$ over the recipient $j$ because the donor can meet anyone as the recipient with the same probability.
Rearranging the terms, we arrive at Eq.~\eqref{eq:motion2}. The factor of $1-2\epsilon_k$ in the summation implies that the assessment rule $\alpha_k$ becomes irrelevant if $\epsilon_k$ is as high as $1/2$. As $t\to\infty$, we assume that the system converges so that $m_{ki}^{t+1} \approx m_{ki}^t$. By this assumption, we can remove $q$ from the equations of motion. Furthermore, if we take into account the symmetry among individuals, we may assume that $m_{i1}^t \approx m_{i2}^t \approx \ldots \approx m_{iN}^t$. For this reason, the summation in Eq.~\eqref{eq:motion2} is approximated as follows:
\begin{eqnarray}
\frac{1}{N} \sum_{j=1}^N \alpha \left[ \beta \left( m_{ij}^t \right), m_{kj}^t \right] &\approx& \frac{1}{N} \alpha\left[ \beta \left( m_{i0}^t \right), m_{k0}^t \right] + \frac{N-1}{N} \alpha\left[ \beta \left( m_{i1}^t \right), m_{k1}^t \right]\\
&\approx& \alpha\left[ \beta \left( m_{i1}^t \right), m_{k1}^t \right].
\end{eqnarray}
Note that we have used the condition that $N \gg 1$ in the second line. Thus,} we obtain the following set of coupled nonlinear equations in a stationary state:
\begin{subequations}
\begin{align}
0 \approx& -m_{00} + \epsilon_0 + \left( 1-2\epsilon_0 \right) \alpha \left[ \beta \left( m_{01} \right), m_{01} \right] \equiv f_1 \left( m_{00}, m_{01} \right)\label{eq:m00}\\
0 \approx& -m_{01} + \epsilon_0 + \left( 1-2\epsilon_0 \right) \alpha \left[ \beta \left( m_{11} \right), m_{01} \right] \equiv f_2 \left( m_{01}, m_{11} \right)\label{eq:m01}\\
0 \approx& -m_{10} + \epsilon_1 + \left( 1-2\epsilon_1 \right) \alpha \left[ \beta \left( m_{01} \right), m_{11} \right] \equiv f_3 \left( m_{01}, m_{10}, m_{11} \right)\label{eq:m10}\\
0 \approx& -m_{11} + \epsilon_1 + \left( 1-2\epsilon_1 \right) \alpha \left[ \beta \left( m_{11} \right), m_{11} \right] \equiv f_4 \left( m_{11} \right), \label{eq:m11}
\end{align}
\label{eq:coupled}
\end{subequations}
where we have again suppressed the superscript $t$. Due to the structure of the problem, we can obtain the four unknowns from $m_{00}$ to $m_{11}$ one by one in the following way. Firstly, we solve Eq.~\eqref{eq:m11} for $m_{11}$. Secondly, we substitute $m_{11}$ in Eq.~\eqref{eq:m01} and solve the equation for $m_{01}$. Then, we substitute $m_{11}$ and $m_{01}$ in Eq.~\eqref{eq:m10} and solve the equation for $m_{10}$. Lastly, we substitute $m_{01}$ in Eq.~\eqref{eq:m00} and solve the equation for $m_{00}$, but this last step can be skipped because $m_{00}$ is irrelevant to the payoff difference in Eq.~\eqref{eq:ss_dpi0}.

\section{Results}
\label{sec:results}

\subsection{Image Scoring and Stern Judging}
\begin{figure}
\includegraphics[width=0.45\columnwidth]{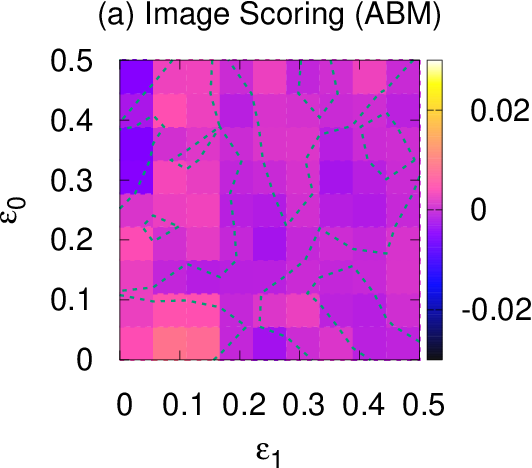}
\includegraphics[width=0.45\columnwidth]{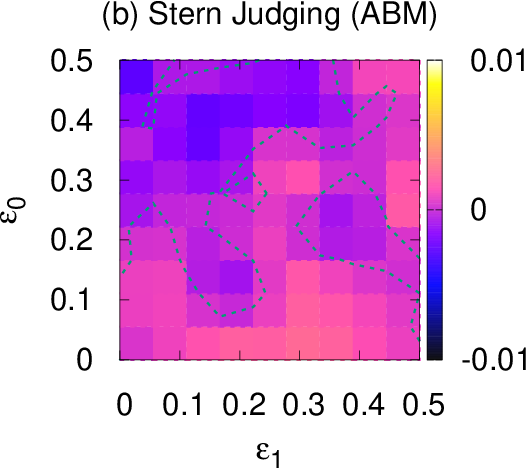}
\caption{A mutant's payoff advantage relative to a resident's, with dotted lines showing the contours of $\Delta\pi_0 = 0$.
We have run ABM simulations with $N=10^2$, $T=10^4$, and $S=10^3$. The donation game is parametrized by $b=1$ and $c=1/2$, and the observation probability is set to $q=1$.
(a) Numerical results for Image Scoring, which is zero everywhere up to statistical noise, in full agreement with the prediction from the continuous model. (b) Stern Judging shows similar behavior as predicted by our analysis.}
\label{fig:scsj}
\end{figure}

In the continuous formulation, Image Scoring is described as $\alpha \left( \beta_{ij}^t, m_{kj}^t \right) = \beta_{ij}^t$, while Stern Judging has $\alpha \left( \beta_{ij}^t, m_{kj}^t \right) = 2\beta_{ij}^t m_{kj}^t - \beta_{ij}^t - m_{kj}^t +1$.
In either case, the result is a perfect neutrality with $\Delta \pi_0 = 0$ because $m_{00} = m_{01} = m_{10} = m_{11} = 1/2$, regardless of $\epsilon_0$ and $\epsilon_1$. This result makes sense: It has been well known that Image Scoring cannot sustain cooperation~\cite{ohtsuki2004should} \skb{(see also Appendix~\ref{app:second} for a duality-based argument)}, and the stochastic dynamics induced by Stern Judging is known to be driven by entropy~\cite{bae2024exact}. Our numerical results confirm this prediction as shown in Fig.~\ref{fig:scsj}.

\subsection{Shunning}
\begin{figure}
\includegraphics[width=0.45\columnwidth]{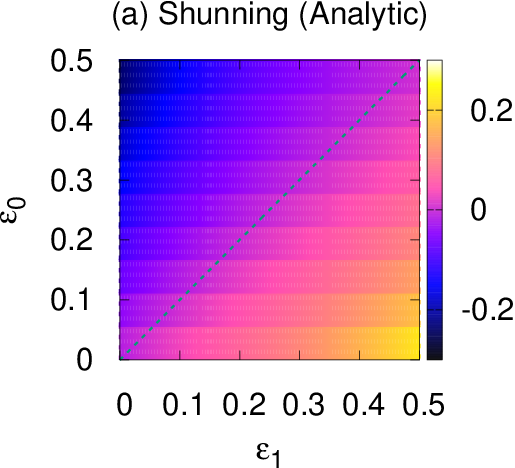}
\includegraphics[width=0.45\columnwidth]{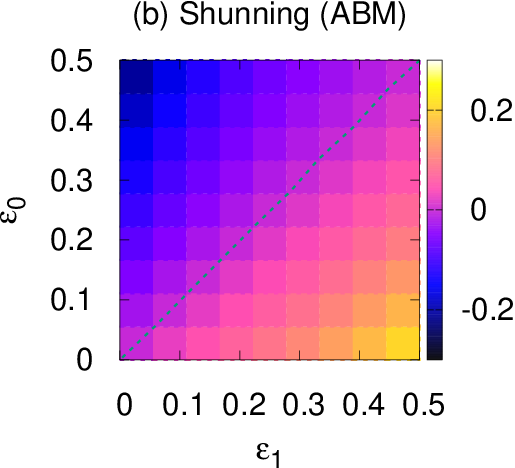}
\caption{A mutant's payoff advantage relative to a resident according to Shunning [see Eq.~\eqref{eq:sh}]. The dotted lines show the contour lines of $\Delta \pi_0 = 0$. (a) Analytic results obtained by solving the continuous model. (b) ABM results with the same simulation parameters as in Fig.~\ref{fig:scsj}.
}
\label{fig:shun}
\end{figure}

We now consider Shunning, whose assessment rule is given as follows:
\begin{equation}
\alpha\left(\beta_{ij}^t, m_{kj}^t \right) = \beta_{ij}^t m_{kj}^t.
\label{eq:sh}
\end{equation}
When we solve Eq.~\eqref{eq:coupled}, the solution is given as
\begin{subequations}
\begin{align}
m_{01} &= \frac{2\epsilon_0 \left(1-2\epsilon_1\right)}{1+2\left(\epsilon_0 - 2\epsilon_1\right) + \left(1-2\epsilon_0\right)\sqrt{1-4\epsilon_1+8\epsilon_1^2}}\\
m_{10} &= \epsilon_1 + \frac{\epsilon_0 \left(1 - 2 \epsilon_1\right) \left(1 - \sqrt{1 - 4 \epsilon_1 + 8 \epsilon_1^2}\right)}{1+2\left(\epsilon_0 - 2\epsilon_1\right) + \left(1-2\epsilon_0\right)\sqrt{1-4\epsilon_1+8\epsilon_1^2}}\\
m_{11} &= \frac{1-\sqrt{1-4\epsilon_1+8\epsilon_1^2}}{2\left( 1-2\epsilon_1 \right)},
\end{align}
\end{subequations}
and it predicts the payoff disadvantage for a mutant with $\epsilon_0 > \epsilon_1$ for any $\epsilon_1$ between $0$ and $1/2$. \skb{The mutation-selection dynamics will favor error-free mutants if the social norm is Shunning. Thus, it is an example of a norm that is resistant to assessment errors.} We plot the analytic result in Fig.~\ref{fig:shun}(a). It agrees well with the ABM result shown in Fig.~\ref{fig:shun}(b).

\subsection{Simple Standing}
\begin{figure}
\includegraphics[width=0.45\columnwidth]{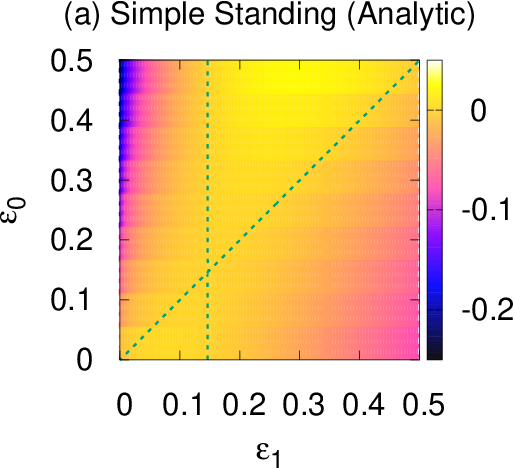}
\includegraphics[width=0.45\columnwidth]{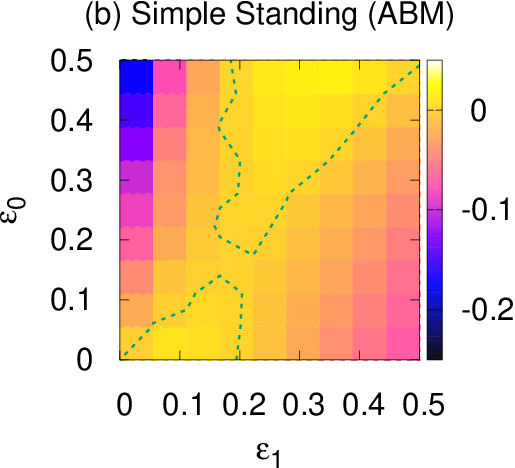}
\caption{Visualization of Eq.~\eqref{eq:ss_dpi0} in the case of Simple Standing, where the dotted lines show the contour lines of $\Delta \pi_0 = 0$. (a) Analytic results from the continuous model. (b) ABM results with the same simulation parameters as in Figs.~\ref{fig:scsj} and \ref{fig:shun}.
}
\label{fig:L3}
\end{figure}

The last norm is Simple Standing, whose assessment rule is given as
$\alpha\left(\beta_{ij}^t, m_{kj}^t \right) = \beta_{ij}^t m_{kj}^t - m_{kj}^t + 1$.
The corresponding continuous model is solved by
\begin{subequations}
\begin{align}
m_{01}^\ast &= \frac{(1-\epsilon_0) (1-2\epsilon_1)}{1-(3-2\epsilon_0)\epsilon_1 + (1-2\epsilon_0) \sqrt{(1-\epsilon_1)\epsilon_1}}\label{eq:m01ss}\\
m_{10}^\ast &= \frac{(1-\epsilon_1)^2 - \epsilon_0 (1-\epsilon_1) + (\epsilon_0 - \epsilon_1) \sqrt{(1-\epsilon_1)\epsilon_1}}{1-(3-2\epsilon_0)\epsilon_1 + (1-2\epsilon_0) \sqrt{(1-\epsilon_1)\epsilon_1}}\label{eq:m10ss}\\
m_{11}^\ast &= \frac{1-\epsilon_1-\sqrt{(1-\epsilon_1)\epsilon_1}}{1-2\epsilon_1},\label{eq:m11ss}
\end{align}
\label{eq:analytic_ss}
\end{subequations}
In Fig.~\ref{fig:L3}(a), we plot the resulting $\Delta \pi_0$ in the $(\epsilon_1, \epsilon_0)$ plane, assuming that $b=1$ and $c=1/2$.
It shows the existence of a threshold: If $\epsilon_1 < \epsilon^\ast$, where
\begin{equation}
\epsilon^\ast \equiv \frac{1}{4} (3-2r-\sqrt{1+4r-4r^2})
\label{eq:threshold}
\end{equation}
with $r\equiv c/b$, mutants more prone to errors with $\epsilon_0 > \epsilon_1$ cannot invade because $\Delta \pi_0 <0$. This calculation is fully corroborated by numerical simulations [Fig.~\ref{fig:L3}(b)]. \skb{The differences from Fig.~\ref{fig:L3}(a) are mainly due to the finite resolution and statistical fluctuations of our ABM simulations.}

\begin{figure}
\includegraphics[width=0.45\columnwidth]{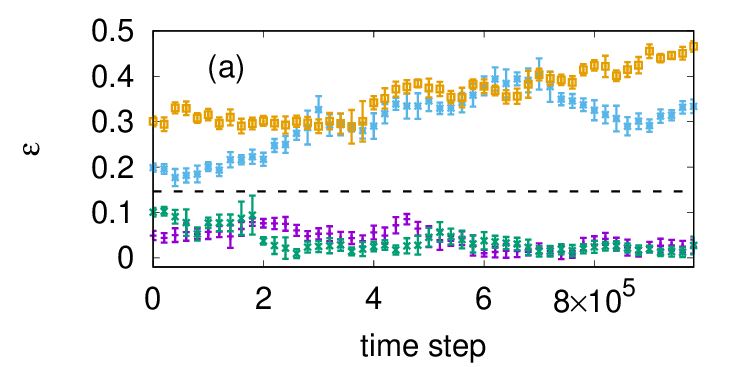}
\includegraphics[width=0.45\columnwidth]{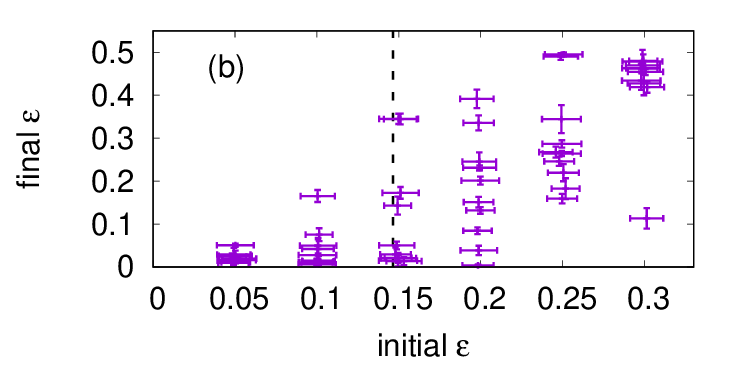}
\caption{\skb{Mutation-selection dynamics with $N=50$ and $r\equiv c/b=1/2$. The total simulation period is $T=10^6$ time steps, and the periods for selection and mutation are $\tau_s=10^3$ and $\tau_m=10^4$, respectively. The mutation is represented by the Gaussian random variable with width $w=10^{-2}$. (a) Sample trajectories starting from mean values $\bar\epsilon = 0.05$, $0.1$, $0.2$, and $0.3$, respectively. The horizontal dotted line indicates the theoretical threshold $\epsilon^\ast =(2-\sqrt{2})/4 \approx 0.15$ [Eq.~\eqref{eq:threshold}].
(b) The distribution of $\left\{ \epsilon_i \right\}$ after the $T=10^6$ time steps, obtained from $10$ sample trajectories for each of the initial mean values. The vertical dotted line indicates $\epsilon^\ast$.}}
\label{fig:additional}
\end{figure}

\skb{To elaborate on the above result, we have also simulated the mutation-selection dynamics of Simple Standing as has been explained in Sec.~\ref{sec:selection}. Figure~\ref{fig:additional}(a) shows a few sample trajectories. The average probability of error tends to decrease if it starts from a low value and vice versa, which coincides with our analytic prediction based on the continuous model.
In Fig.~\ref{fig:additional}(b), we have generated $10$ samples with $N=50$ and $r=1/2$ for each starting point and checked where each sample ends up after $T=10^6$ time steps. Although the finite-size effect may induce some difference between the predicted threshold and the effective one in ABM simulations, we observe a clear difference in the long-term behavior of the error probabilities depending on whether the initial value is below or above a certain threshold.
}

\section{Discussion and Summary}
\label{sec:summary}

\begin{figure}
\includegraphics[width=0.45\columnwidth]{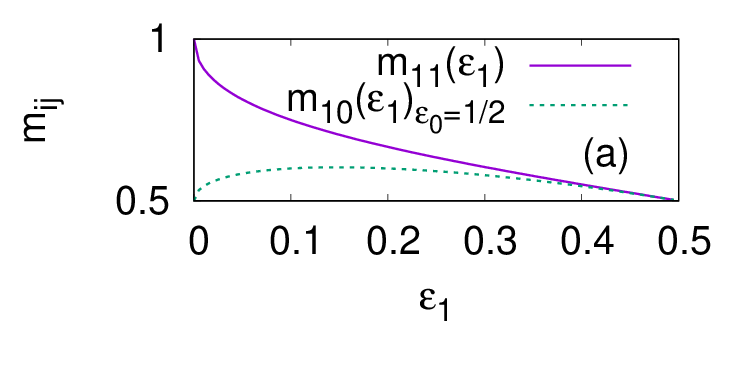}
\includegraphics[width=0.45\columnwidth]{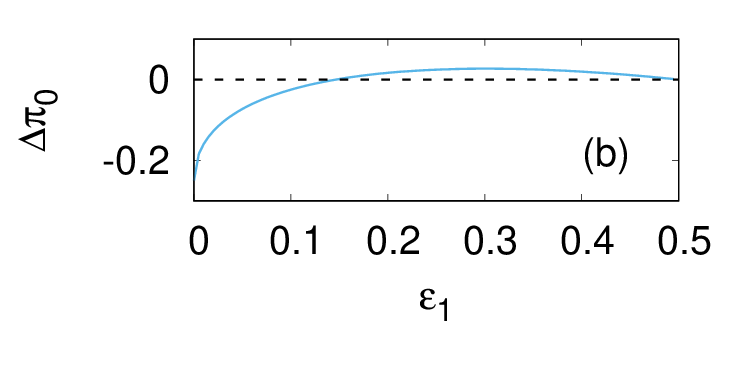}
\caption{(a) Equations~\eqref{eq:m10ss} and \eqref{eq:m11ss} for Simple Standing, when $\epsilon_0=1/2$, at which $m_{01}$ is identically $1/2$ [Eq.~\eqref{eq:m01ss}]. (b) The resulting $\Delta \pi_0$ [Eq.~\eqref{eq:ss_dpi0}] for Simple Standing, when $b=1$ and $c=1/2$. The horizontal dashed line is drawn to show where $\Delta \pi_0$ changes the sign.}
\label{fig:ssrand}
\end{figure}

In studying the effects of assessment errors, we often assume that errors occur with a low probability~\cite{lee2021local,fujimoto2022reputation}.
Then, if Simple Standing governs society, most people regard each other as good, and even if someone is misjudged as bad, the error will be corrected as soon as the misunderstood person properly follows the norm. Our analysis shows that this is the regime below the threshold, where error proneness is simply taken as a sign of defection, and hence suppressed. However, if the probability of error exceeds the threshold, it turns out that the cooperativeness of Simple Standing opens the door to error-prone individuals.

To understand how it happens, imagine that $\epsilon_1$ gradually decreases from $1/2$, while $\epsilon_0$ is fixed at $1/2$ [Fig.~\ref{fig:ssrand}(a)].
When $\epsilon_1=1/2$, the assessment rule is meaningless, and everyone looks half good to each other. As $\epsilon_1$ decreases, Simple Standing begins to work and $m_{11}$ increases above $1/2$. More importantly, $m_{10}$ also increases with the same slope. Residents can clearly tell the difference between mutants and themselves when $\epsilon_1$ is sufficiently low, so $m_{10}$ converges to $1/2$ as $\epsilon_1 \to 0$, while $m_{11}$ converges to $1$. However, before reaching such a low probability of errors, a resident player would donate more to the mutant than the mutant does to the resident player in return. The result is $\Delta \pi_0>0$ for $\epsilon_1 > \epsilon_1^\ast$ [Fig.~\ref{fig:ssrand}(b)].

Now, if mutants have $\epsilon_0 \le 1/2$ in general, it is true that their difference is recognized by the resident population, so that $m_{10}$ remains lower than $m_{11}$ as long as $\epsilon_0 > \epsilon_1$. However, due to the structure of the social dilemma, defection pays better than cooperation, making careless assessments by mutants more favorable even if $\epsilon_0 > \epsilon_1$, provided that $\epsilon_1$ is above the threshold $\epsilon^\ast$. The threshold $\epsilon^\ast$ is a decreasing function of $r \equiv c/b$, and it is straightforward to see from Eq.~\eqref{eq:threshold} that $\epsilon^\ast(r=0) = 1/2$ and $\epsilon^\ast(r=1) = 0$. In Fig.~\ref{fig:L3} where $r=1/2$, the threshold is already as low as $\epsilon^\ast = \left(2-\sqrt{2}\right)/4 \approx 0.15$. If the probability of assessment errors increases higher than $\epsilon^\ast$ in the population, any individual effort to acquire precise information about others will be futile because the result should be worse than a coin toss. The social norm breaks down.

In summary, we have suggested when a social norm can become unstable in the presence of randomness in assessments. The crucial assumption is that the probability of assessment errors is not an entirely environmental factor common to everyone, but at least partly an individual attribute, which may in principle differ individual by individual in their tendencies to pay attention to others and observe the social norm carefully. After examining the four well-known second-order norms, we have presented Simple Standing as an example of threshold behavior, which makes the resident population vulnerable to mutants with higher error probabilities when its own $\epsilon_1$ exceeds a threshold. This behavior qualitatively explains why Simple Standing rewards error-prone individuals when the probabilities of assessment errors are heterogeneously distributed between low and high values in the population~\cite{le2025response}.

Before concluding this work, we add the following words of caution:
Despite the success of the continuous model applied to the four second-order norms, it does not always reproduce the results of the corresponding discrete model. In particular, the case of Staying (also known as L7 in the context of the leading eight) is instructive. Here again, the continuous model predicts $m_{00}^\ast = m_{01}^\ast = m_{10}^\ast = m_{11}^\ast = 1/2$ regardless of $\epsilon_0$ and $\epsilon_1$. However, this contradicts the stochastic ABM dynamics of Staying, which results in $m_{ij}=1$ for any $i$ and $j$, as long as the probability of error is low enough~\cite{fujimoto2024leader,bae2025indirect}. This clear mismatch calls for a more systematic analysis of higher-order norms \skb{that better captures the probabilistic nature of the dynamics}.

\begin{acknowledgments}
We acknowledge support by Basic Science Research Program through the National Research Foundation of Korea (NRF) funded by the Ministry of Education (NRF-2020R1I1A2071670).
\end{acknowledgments}

%

\newpage
\appendix

\setcounter{table}{0}
\renewcommand{\thetable}{A\arabic{table}}

\setcounter{figure}{0}
\renewcommand{\thefigure}{A\arabic{figure}}

\section{\skb{Second-order norms}}
\label{app:second}

\begin{figure}
\includegraphics[width=0.45\textwidth]{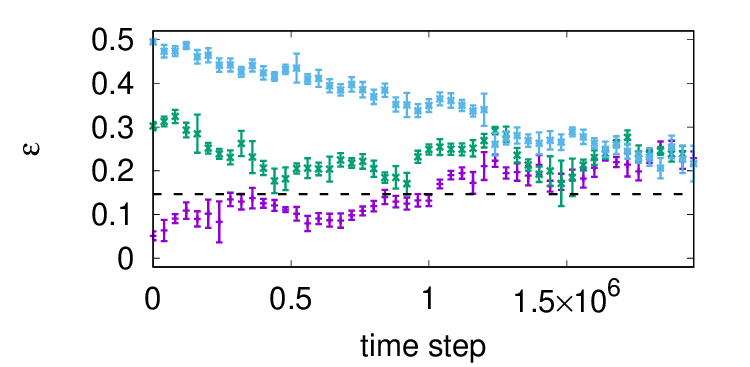}
\caption{\skb{Sample trajectories of $S_{13}$, the dual norm of Simple Standing, for $T=2\times 10^6$ time steps. The other simulation parameters are the same as in Fig.~\ref{fig:additional}(a), and the dotted line again shows $\epsilon = (2-\sqrt{2})/4 \approx 0.15$. The effective point of the mutation-selection equilibrium for this small system exists around $0.2$, consistently with Fig.~\ref{fig:additional}(b).}}
\label{fig:dual}
\end{figure}

\begin{figure}
\includegraphics[width=0.45\columnwidth]{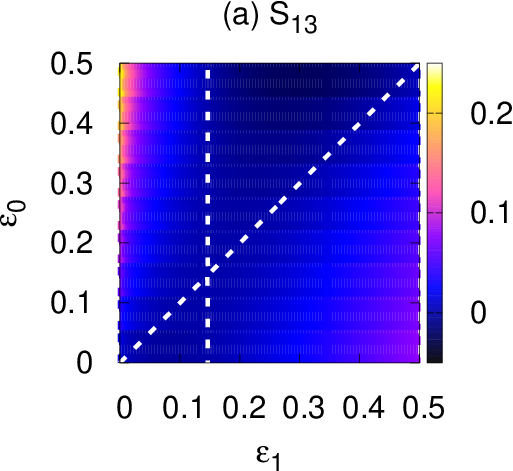}
\includegraphics[width=0.45\columnwidth]{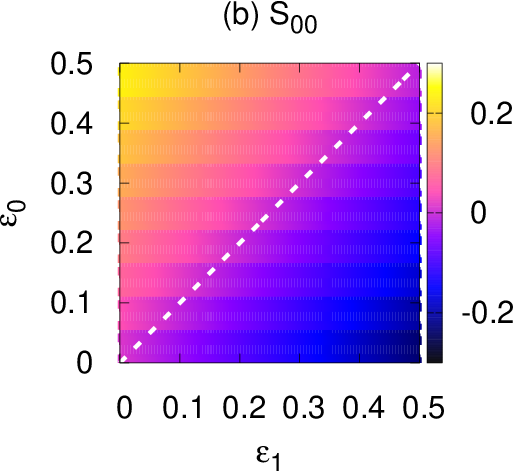}\\
\includegraphics[width=0.45\columnwidth]{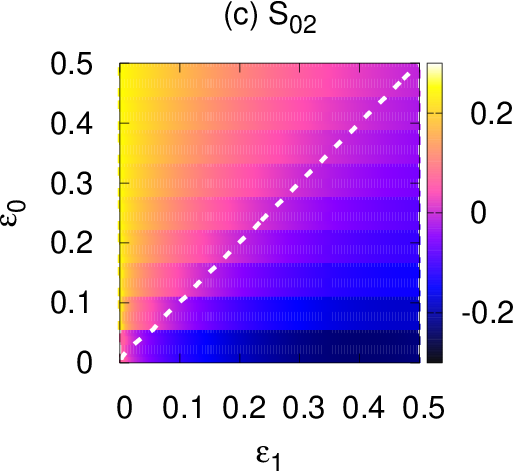}
\includegraphics[width=0.45\columnwidth]{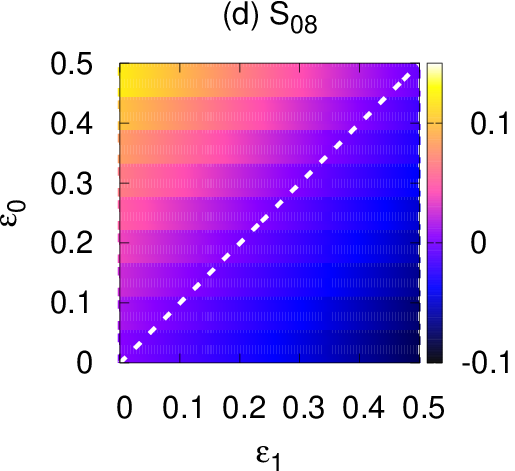}
\caption{\skb{Visualization of $\Delta \pi_0$ [Eq.~\eqref{eq:ss_dpi0}] for four norms in Table~\ref{tab:second}. The dotted lines show the contour lines of $\Delta \pi_0 = 0$.}}
\label{fig:others}
\end{figure}

\skb{In this appendix, we will only present calculations based on the continuous model, but they are all reproduced by ABM simulations of the original discrete model.
In Table~\ref{tab:second}, we have listed all $16$ possible assessment rules of second-order norms, where $\beta(G)=C$ and $\beta(B)=D$ are fixed. Only Simple Standing shows the threshold behavior.
Before verifying this statement, let us point out that the norms are related by a duality relation. Let us say that a norm $S_\mu$ is dual to $S_\nu$ if we obtain $S_\nu$ by changing $C$ to $D$ and $G$ to $B$ in $S_\mu$. For example, $S_{13}$ is dual to Simple Standing ($S_{04}$) because they are related in the following way:
\begin{subequations}
\begin{align}
\alpha_{04}(C,G) = G &\longleftrightarrow \alpha_{13}(D,B) = B\\
\alpha_{04}(D,G) = B &\longleftrightarrow \alpha_{13}(C,B) = G\\
\alpha_{04}(C,B) = G &\longleftrightarrow \alpha_{13}(D,G) = B\\
\alpha_{04}(D,B) = G &\longleftrightarrow \alpha_{13}(C,G) = B,
\end{align}
\end{subequations}
where the subscripts indicate the corresponding social norms between $S_{04}$ and $S_{13}$. If we obtain the same after this duality transformation, we will say that the norm is self-dual.
Assume that $S_\mu$ and $S_\nu$ are dual to each other. If Eq.~\eqref{eq:coupled} is solved by $\left(m_{00,\mu}^\ast, m_{01,\mu}^\ast, m_{10,\mu}^\ast, m_{11,\mu}^\ast \right)$ and $\left(m_{00,\nu}^\ast, m_{01,\nu}^\ast, m_{10,\nu}^\ast, m_{11,\nu}^\ast \right)$ for $S_\mu$ and $S_\nu$, respectively, some algebra shows the following equalities:
\begin{equation}
m_{00,\mu}^\ast + m_{00,\nu}^\ast =
m_{01,\mu}^\ast + m_{01,\nu}^\ast =
m_{10,\mu}^\ast + m_{10,\nu}^\ast =
m_{11,\mu}^\ast + m_{11,\nu}^\ast = 1.
\end{equation}
For a self-dual norm, it is obvious that $m_{00}^\ast = m_{01}^\ast = m_{10}^\ast = m_{11}^\ast = 1/2$. Another immediate consequence is that the payoff advantage $\Delta \pi_0$ of a mutant [Eq.~\eqref{eq:ss_dpi0}] has the opposite sign after the duality transformation.
As a result, $\epsilon_1 = (2-\sqrt{2})/4 \approx 0.15$ for $S_{13}$ will no longer be a threshold, but a point around which the probability of assessment error fluctuates through mutation and selection [Figs.~\ref{fig:dual} and \ref{fig:others}(a)].
}
\begin{table}
\caption{\skb{Second-order norms. The behavioral rule is fixed as $\beta(G)=C$ and $\beta(B)=D$.}}
\begin{ruledtabular}
\begin{tabular}{c|cccc|c}
 & $\alpha(C,G)$ & $\alpha(D,G)$ & $\alpha(C,B)$ & $\alpha(D,B)$ & misc.\\\hline
$S_{00}$ & $G$ & $G$ & $G$ & $G$ & Eq.~\eqref{eq:s00}\\
$S_{01}$ & $G$ & $G$ & $G$ & $B$ & dual to Shunning\\
$S_{02}$ & $G$ & $G$ & $B$ & $G$ & Eq.~\eqref{eq:s02}\\
$S_{03}$ & $G$ & $G$ & $B$ & $B$ & self-dual\\
$S_{04}$ & $G$ & $B$ & $G$ & $G$ & Simple Standing\\
$S_{05}$ & $G$ & $B$ & $G$ & $B$ & Image Scoring (self-dual)\\
$S_{06}$ & $G$ & $B$ & $B$ & $G$ & Stern Judging\\
$S_{07}$ & $G$ & $B$ & $B$ & $B$ & Shunning\\
$S_{08}$ & $B$ & $G$ & $G$ & $G$ & Eq.~\eqref{eq:s08}\\
$S_{09}$ & $B$ & $G$ & $G$ & $B$ & dual to Stern Judging\\
$S_{10}$ & $B$ & $G$ & $B$ & $G$ & self-dual\\
$S_{11}$ & $B$ & $G$ & $B$ & $B$ & dual to $S_{02}$\\
$S_{12}$ & $B$ & $B$ & $G$ & $G$ & self-dual\\
$S_{13}$ & $B$ & $B$ & $G$ & $B$ & dual to Simple Standing\\
$S_{14}$ & $B$ & $B$ & $B$ & $G$ & dual to $S_{08}$\\
$S_{15}$ & $B$ & $B$ & $B$ & $B$ & dual to $S_{00}$
\end{tabular}
\label{tab:second}
\end{ruledtabular}
\end{table}

\skb{Table~\ref{tab:second} has four self-dual norms, one of which is Image Scoring. In the main text, we have also obtained results for Stern Judging, Shunning, and Simple Standing. Thus, we need to consider only $S_{00}$, $S_{02}$, and $S_{08}$ here. The first one, $S_{00}$, regards everything as good unless error occurs. The solution of Eq.~\eqref{eq:coupled} is thus obtained as follows:
\begin{subequations}
\begin{align}
m_{01}^\ast&=1-\epsilon_0\\
m_{10}^\ast&=1-\epsilon_1\\
m_{11}^\ast&=1-\epsilon_1.
\end{align}
\label{eq:s00}
\end{subequations}
Clearly, this does not show any threshold behavior [Fig.~\ref{fig:others}(b)].
As for $S_{08}$, the solution looks more complicated as written below,
\begin{subequations}
\begin{align}
m_{01}^\ast&=\frac{\epsilon_1-2\epsilon_0^2-\epsilon_0(1-2\epsilon_0) \left[ \epsilon_1 + \sqrt{\epsilon_1 (1-\epsilon_1)} \right]}{\epsilon_1-4\epsilon_0^2(1-\epsilon_1)}\\
m_{10}^\ast&=\frac{\epsilon_1 +\epsilon_0 \epsilon_1 (1-2\epsilon_1)-4\epsilon_0^2(1-\epsilon_1)-(\epsilon_1-2\epsilon_0^2)\sqrt{\epsilon_1 (1-\epsilon_1)}}{\epsilon_1-4\epsilon_0^2(1-\epsilon_1)}\\
m_{11}^\ast&=\frac{1-\epsilon_1-\sqrt{\epsilon_1 (1-\epsilon_1)}}{1-2\epsilon_1},
\end{align}
\label{eq:s02}
\end{subequations}
but the overall behavior is still monotone [Fig.~\ref{fig:others}(c)]. The same applies to $S_{08}$. We solve Eq.~\eqref{eq:coupled} as follows:
\begin{subequations}
\begin{align}
m_{01}^\ast&=\frac{2(1-\epsilon_0)(1-2\epsilon_1)}{1+2\epsilon_0-4\epsilon_1+(1-2\epsilon_0)\sqrt{5-12\epsilon_1+8\epsilon_1^2}}\\
m_{10}^\ast&=\frac{2+\epsilon_0-7\epsilon_1+4\epsilon_1^2+(\epsilon_1-\epsilon_0)\sqrt{5-12\epsilon_1+8\epsilon_1^2}}{1+2\epsilon_0-4\epsilon_1+(1-2\epsilon_0)\sqrt{5-12\epsilon_1+8\epsilon_1^2}}\\
m_{11}^\ast&=\frac{\sqrt{5-12\epsilon_1+8\epsilon_1^2}-1}{2(1-2\epsilon_1)},
\end{align}
\label{eq:s08}
\end{subequations}
and this solution does not show any threshold behavior [Fig.~\ref{fig:others}(d)].
}

\end{document}